# Programming the Universe: The First Commandment of Software Engineering for all Varieties of Information Systems[0,1,2]

Silvio Meira[ι,ω], Vanilson Burégio[σ], Paulo Borba[ι],
Vinicius Garcia[ι], Jones Albuquerque[σ,κ], Sergio Soares[ι,τ]

[ι]Informatics Center, UFPE
[ω]TDS.company
[σ]Statistics and Informatics Department, UFRPE
[κ]Keizo Asami Laboratory of Imunopathology, UFPE
[τ]SENAI Innovation Institute for Informatics, ISI-TICs
{srlm, phmb, vcg, scbs}@cin.ufpe.br
vanilson.buregio@ufrpe.br, jones.albuquerque@pq.cnpq.br

## ABSTRACT

Since the early days of computers and programs, the process and outcomes of software development has been a minefield plagued with problems and failures, as much as the complexity and complication of software and its development has increased by a thousandfold in half a century. Over the years, a number of theories, laws, best practices, manifestos and methodologies have emerged, with varied degrees of (un)success. Our experience as software engineers of complex and large-scale systems shows that those guidelines are bound to previously defined and often narrow scopes. Enough is enough. Nowadays, nearly every company is in the software and services business and everything is - or is managed by - software. It is about time, then, that the laws that govern our universe ought to be redefined. In this context, we discuss and present a set of universal laws that leads us to propose **the first commandment of software engineering** for all varieties of information systems.

---

[0] ***A version of this paper was accepted for presentation and publication at XXX SBES, 30th Brazilian Symposium on Software Engineering,*** *in the* ***Insightful Ideas*** *track; the event took place in Maringá-PR, Brazil, in September 2016.*

[1] This work should be cited as *Silvio Meira, Vanilson Burégio, Paulo Borba, Vinicius Garcia, Jones Albuquerque, and Sergio Soares. 2016.* ***Programming the Universe: The First Commandment of Software Engineering for all Varieties of Information Systems****. In Proceedings of the 30th Brazilian Symposium on Software Engineering (SBES '16), Eduardo Santana de Almeida (Ed.). ACM, New York, NY, USA, 153-156. DOI: http://dx.doi.org/10.1145/2973839.2982567.*

[2] The research that resulted in this paper was partially supported by the National Institute of Science and Technology for Software Engineering (INES.org.br -of which all authors are members), funded by CNPq grant #573964/2008-4.

## 1. INTRODUCTION

At a time when some of the world's top scientific minds[1] believe that the universe including all things, inanimate or alive, are not only based on code but are indeed code [3, 14], running on apparently different platforms but, all in all, most probably based upon the same basic set of universal constructs, we can't avoid starting to think things of all sorts as information systems.

According to Deutsch [1], *Every finitely realizable physical system can be perfectly simulated by a universal model computing machine operating by finite means*, which can be regarded as the extension of the Strong Church-Turing Thesis [16] for the physical world. Reusing Bostrom's *simulation argument* [2], we can state that *every sufficiently advanced civilization will develop the capability of software writing* and that *such capability will be used both for 1. Simulating significant parts of its context in software* and, as soon as it can, and with much wider impact and consequences, *2. (Re)Writing the software of its context*.

Indeed, we have just started to deal with ***life as code*** and are facing the initial problems of (re)programming bacteria [3] for our own means and ends and, on the other hand, trying to create life from scratch and searching for its basic building blocks [8], the ***combinators*** of living things.

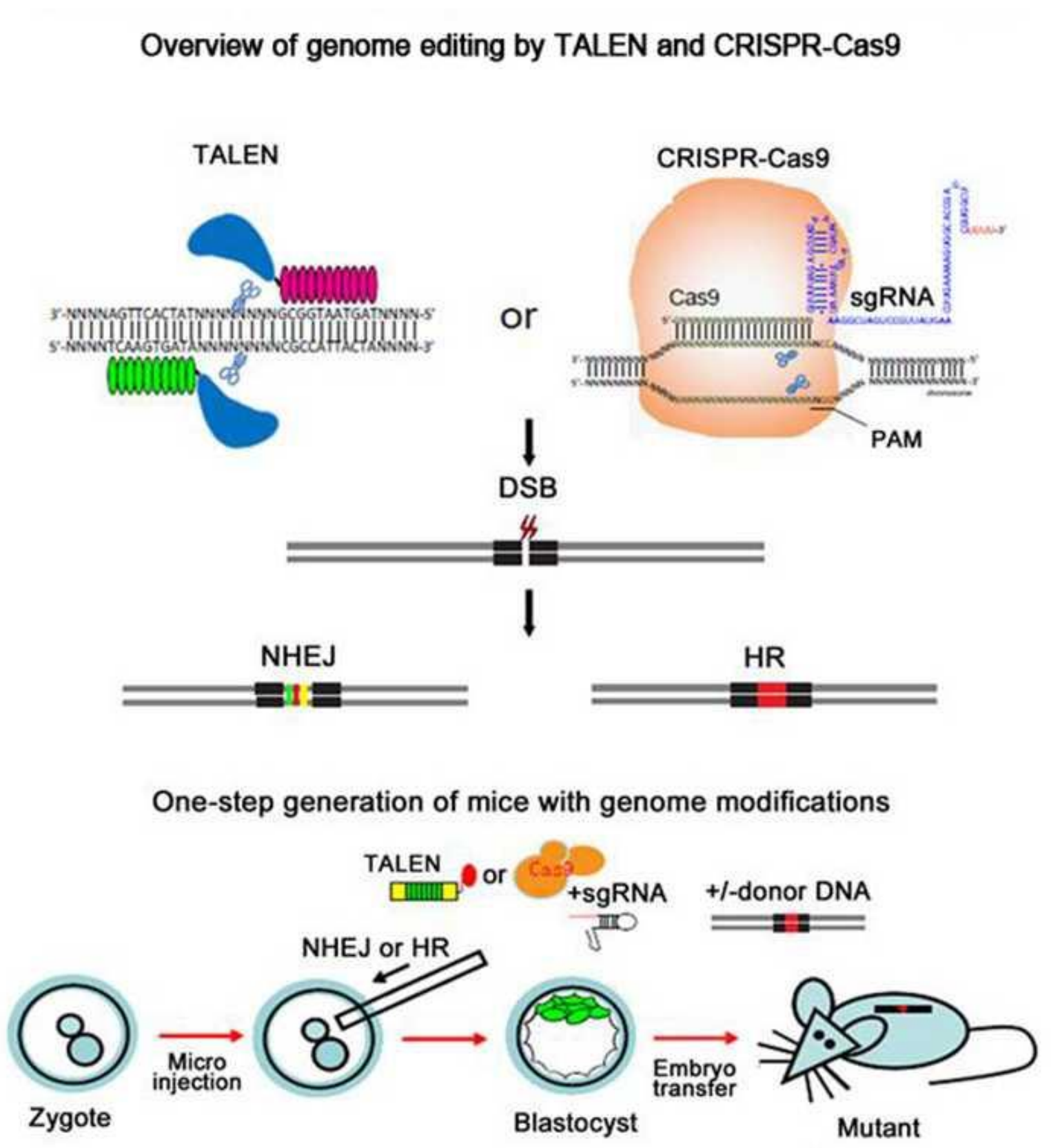

**Figure 1 Genome editing process using CRISPR [13].**

[1] The 17th annual Isaac Asimov Debate (April 12, 2016) at New York's American Museum of Natural History sold out in just 3 minutes online, host Neil deGrasse Tyson told the audience. The debate featured five experts chewing on the idea of the universe as a simulation. Discover more in the link http://bit.ly/1VUNIrL.

After more than forty years of experience with software engineering for writing programs for computers, most of it associated to the development of corporate information systems (with a fair to high degree of failure) and in a minor degree to industrial and embedded systems (also with its problems and associated learning), what can we say to the newcomers, those that will ***edit*** (to remove and add features), ***fix*** (to correct traits) or ***write*** (to create new "stuff") DNA, as recently (10 march 2016) proposed in [9] and illustrated in a real process in Figure 1, for example?

Are there any ***laws***, derived from our long experience, that are indeed ***universal*** and that we could teach others, and in other areas of ***coding***, so that they depart several degrees above the point where we started, in a sort of universal logarithmic scale of ***software development competence***? Is there such scale? Are there universally valid ***software engineering laws***?

If there are, it is worth the while to initiate a search for such nuggets of quasi-theoretical software engineering knowledge to share with others, because we are on the verge of seeing software becoming an almost universal trait of all systems, in all areas of human knowledge and expertise [10]. Otherwise, a rerun of the long series of problems we -software engineers- had since our software systems were written as circuits will be seen, because -for example- biological software is threading the same path, including using the same languages, like Verilog [3], as shown in Figure 2.

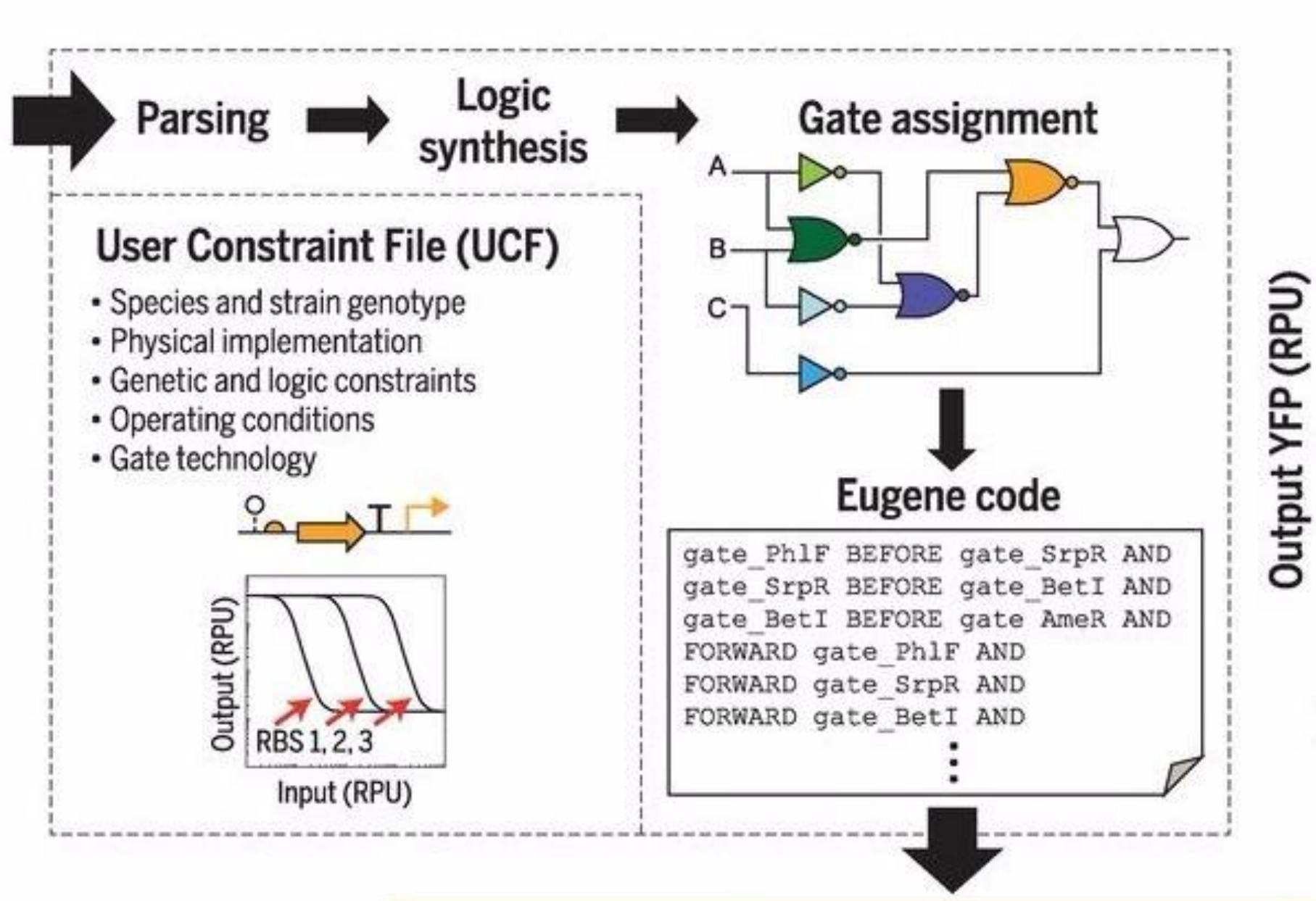

**Figure 2 A user specifies the desired circuit function in Verilog code, and this is transformed into a DNA sequence [3].**

If there aren't, it would be a great endeavor to discover whether this is only a ***temporary*** -having no laws so far[2], but good reasons to believe we will have some, universal, sometime in the future- or ***definitive*** condition, in which we ought to state why there are not and there will be no universal software engineering laws whatsoever, whenever.

---

[2] More on that later, when we refer to Lehman's 7th Law.

And that will mean, of course, that we are condemned to a future of hacking, *spaghetting*, and all of its consequences, to life, the universe and everything -and, of course, to the meaning of it all.

Better not be so, then.

This effort tries to state and justify a basic set of **universal** (in the sense of *for everything, everywhere, all the time*) **software** (in the sense of *code*, any kind of) **engineering** (in the sense of understanding *what* things, concrete and abstract, *are*; *what* are they *for*; *how* do they *function*; and *how much* they *cost*) ***laws*** (in the sense of fundamental, widely accepted rules of development and behavior of things-in-a-system). In the context of a *Theory for Design and Action* [4], where *laws are statements that say how something should be done in practice*.

We call this first attempt **SELFIE**, an acronym for ***Software Engineering Laws For Information [systems in] Everything***. In this paper, some of the laws are going to be fused into a *commandment*, only a few of which are needed, in the standard interpretation, to build a *culture* upon. To state, in a *concise* and *precise* way, the fundamental laws (***Das Gesetz***) of a universal software culture is the ultimate aim of this effort. This work treats the *first* of a series of commandments only; in due course of time, and incrementally, we will deliver a coherent set of commandments that can serve as basis for our profession, career and community.

Before getting into the real business of this work, maybe it is necessary to reinforce the notion that ***commandments*** need no justification at all; either you believe and follow them, or not. If you believe and obey, salvation through regeneration is guaranteed, in a future that may or may not exist. If you don't, you have to be aware that the (high) cost of your disobedience will very likely be paid in ***this life*** [5].

## 2. SOME FUNDAMENTAL LAWS

We should have at least two sets of laws: one for ***small systems*** (a.k.a. *programs*) and another for ***big systems*** (i.e. *systems themselves*). An intersection of those are the ***universal laws*** that apply to all kinds of systems.

It is very likely that these laws should have nothing to do with specific bits and pieces of programming and architecture, for example, if we want them to be really generic and generally useful.

Take these two examples of desired universal laws of software engineering:

> ***FU****:* *every system, big and small, should have provisions that allow it to be* ***FIXED*** *and* ***UPDATED*** *either from within or outside the system proper;*
>
> ***CK****:* *every system, big and small, should have provisions that allow it to acquire and make use of* ***CONTEXTUAL KNOWLEDGE*** *to decide whether it is functioning properly or not.*

The combination of the two laws gives us the ***FUCK*** property:

> *A system that obeys both the **FU** and **CK** laws can verify the conditions of its operating environment and decide whether its actions are coherent or not with the current context of that environment. Furthermore, if it decides they are not, the system can fix itself, ask for an external fix or be fixed, forcibly, by some sort of external agent.*

Think cells and cellular systems. They are obviously a ***FUCK*** system. And they are extremely resilient under a very broad range of conditions [6]. They've got to obey some kind of lawful combination as above [11].

It should be obvious that Internet of Things (**IoT**, [15]) devices and systems that function properly ought to obey the ***FUCK*** property, otherwise it is quite likely that some funny business might arise from their behavior not being defined by the two laws that lead to it.

We can extend the ***FUCK*** property by stating two further laws:

*IN:* *every system, big and small, should have provisions that allow it to be **INDEPENDENT** of its **NETWORK** as much as possible; in the absence of the surrounding network, systems should gracefully degrade to a level of usefulness that would resemble a fully independent, smaller, isolated system.*

*G:* *in **GENERAL**, for all kinds of systems, **SECURITY** and **INTEGRITY** concerns should supersede those of **FUNCTIONALITY**, which in turn ought to be balanced against **USABILITY**.*

The ***FUCKING*** property is defined by the combination of the above four laws:

> *A system that obeys both the **FU**, **CK**, **IN**, and **G** laws is almost always networked, but will have some acceptable performance when isolated; it can verify the conditions of its operating environment and decide whether its actions are coherent with the current context of that environment. Furthermore, if it decides they are not, the system can fix itself, ask for an external fix or be fixed, forcibly, by some sort of external agent. In all cases, its security, integrity, functionality and usability performances are properly balanced.*

Whatever set of prescriptions we are going to define for software systems, they have to be true and useful in the ***real world***, whether concrete or abstract. Hence the ***REAL*** set of laws, which ought to be valid for all systems, big and small, defined as:

***R****:* *Every system should have provisions that allow it to be **REUSED** in different contexts and by different systems.*

***E****:* *Every system should have provisions that allow it to be **EXTENDED** to accommodate new behaviors.*

> *A:* *Every system should have provisions that allow it to offer* ***ANALYTICS*** *to enable measuring and tracking of its use.*
>
> *L:* *Every system should have provisions that allow it to be* ***LOOSELY COUPLED*** *to reduce both the interdependencies across its modules/components as well as other systems that it possibly interacts with.*

The combination of the above, ***REAL*** properties, with the aforementioned ***FUCKING*** ones leads us to ***REAL FUCKING LAW***, which is the foundation for the development of ***REAL FUCKING systems***.

## 3. THE FIRST COMMANDMENT

The ***REAL FUCKING*** combination of laws is close to what, in religious tradition, is a ***commandment***. If so wanted, one could write it as the ***first commandment*** of software engineering, as we do now:

**Thou shall ONLY develop and deploy REAL FUCKING systems.**

We prefer to write the commandment in a ***positive*** sense, remembering that most commandments are written in a ***negative*** sense. To add an aura of godly respectability to this fundamental software engineering commandment, we would have, in classic hebrew…

**קיים הלא לעולם תזרע ולא תעשה לא**[3,4].

In order to understand the depth and breadth of this commandment, the attentive reader might have already noticed that Lehman's 7th Law of Software Evolution [7]…

> *...the quality of an E-type system will appear to be declining unless it is rigorously maintained and adapted to operational environment changes.*

...is just a ***fact*** that can be ***observed*** in systems for which the first ***commandment*** has been ***disobeyed***. Bad luck. Proper observation of the laws embedded in the ***first*** and the due process deriving from their use will avoid the pitfalls that mar current information system development practice.

---

[3] Pronounced as ***Lo Taaseh vê lo tezrah má she lo kaiam.***

[4] In that old Romance language, Portuguese, ***Você só deve desenvolver e implantar sistemas REALMENTE FODA***.

## 4. CONCLUSION

There is no hope that the ***first commandment*** is going to be uncontroversial and universally accepted. Software engineering is ***engineering*** and not ***science***, even less a formal science. The proof of the laws is going to be their successful use in practice. People will still write software against the set of laws embodied in the ***first commandment***. If those engineers are successful, both in the process of software development and in the resulting information system, the laws and the commandment will of course be invalid.

On the other hand, if those that obey the ***commandment*** and their processes and software are much more successful than those that don't, the ***first*** and other commandments will be the permanent foundations of a new, higher quality, software (development) culture.

We will keep working on the lines that framed this effort to build a set of commandments that will define, once and for all, what software and information systems ought to be and how they should be developed, deployed, used, maintained and, if at all, terminated and with what -and for whom- consequences.

We do think that software is getting out of the realm of computer science and out into the wider, wilder, much more challenging world. As we have already said, all areas of knowledge are bound to be subareas of computer science, programming and, by consequence, subjected to the engineering of their software. The complexity and complications of such are impossible to be dealt with by the naïve theories, methods, processes, environments and tools at our disposal now, even after half a century of what we use to describe as software engineering [12].

In the very near future, ***software liability*** [5] will be a real challenge to every coder and company. If a common and well understood set of laws is not accepted and adopted by all sides in the software ecosystem, which is not less than society as a whole, software disputes, even wars, will plague the very advancement of humanity, given its dependence upon software and its development. And worse will happen if others, and not us, software engineers, lead the process of writing the fundamental laws and commandments of our trade for, supposedly, it us who have the better knowledge to do so. Writing laws and commandments of software engineering that can be understood and accepted by society as a whole then is our job, we can't refrain from trying to do so, and we have to do it as soon as we possibly can, before unlearned adventurers discover and fill the empty space we are leaving to them so far.

We feel we started late in the endeavor to propose a set of commandments for software and its engineering, of which that announced in this paper is just the ***first*** of seven. But it is not too late. Yet. Maybe. Who knows… we hope not.

## 5. ACKNOWLEDGMENTS

We are indebted to Ismar Kaufmann for his precious help with the hebrew translation.